\documentclass{PoS}

\usepackage{amsmath}

\def\to{\rightarrow} 
 
\newcommand\as{\alpha_{\rm S}} 

\def\nn{\nonumber}
\def\ep{\epsilon}
\def\eps{\epsilon}
\def\mb{\mathbf}
\def\mc{\mathcal}
\def\dumunu{d^{\mu\nu}(p;n)}
\def\ddmunu{d_{\mu\nu}(p;n)}

\def\bqt{\mb{q_T}}
\def\qt{$q_T$}
\def\bb{\mb{b}}

\def\bcp{{\boldsymbol{\cal P}}}
\def\bcf{{\boldsymbol{\cal F}}}

\def\cp{{\cal P}}
\def\cf{{\cal F}}

\title{Azimuthally-correlated contributions to QCD transverse-momentum resummation at $\mathcal{O}(\alpha_{\rm S}^2)$}

\ShortTitle{Transverse-momentum dependent collinear functions}
        
\author{Prasanna K. Dhani\\
             INFN, Sezione di Genova,Via Dodecaneso 33, 16146, Genova, Italy\\
             E-mail: \email{prasanna.dhani@ge.infn.it}}
\abstract{Singular factors originating from the QCD factorisation of scattering amplitudes in soft and collinear limits play a prominent role in both organising and computing high-order perturbative contributions to hard-scattering cross sections. In this talk, we start from the factorisation structure of scattering amplitudes in the collinear limit, and we introduce collinear functions that have a process-independent structure. These collinear functions, which are defined at the fully-differential level, can then be integrated over the appropriate observable-dependent phase space to compute logarithmically-enhanced contributions to the corresponding observable. For transverse-momentum dependent observables, we show how the collinear functions can be defined without introducing what is known as rapidity divergences in the literature. We present the results of explicit computations of the collinear functions up to next-to-next-to-leading order in the QCD perturbation theory. }

\FullConference{Loops and Legs in Quantum Field Theory\\ 
                           16th Workshop on Elementary Particle Physics\\
                           April 25-30, 2022\\
		          Ettal, Germany}

\begin{document}
\section{Introduction}
\label{intro}
Transverse-momentum ($\bqt$) resummation is fully developed for the inclusive-production processes of high-mass colourless systems in hadron--hadron collisions. In the kinematical region where the \qt\ of the produced system is much smaller than its invariant mass $M$, the perturbative QCD computation of the \qt-differential cross section leads to large logarithmic contributions of the type $\ln^n (M^2/q_T^2)$. The QCD resummation procedure organises and systematically sums these large contributions to all perturbative orders in the strong coupling $\as$. In this talk we refer to the \qt\ resummation formalism of Refs.~\cite{Collins:1984kg, Catani:2000vq, Catani:2010pd, Catani:2013tia}.

Transverse-momentum resummation is conveniently carried out in impact parameter ($\bb$) space, Fourier conjugate to $\bqt$ space. We directly consider and refer to the notation in Ref.~\cite{Catani:2013tia}. The $\bb$ space cross section at 
$bM \gg 1$ is expressed in terms of the parton distribution functions (PDFs) of the colliding hadrons and of perturbatively calculable factors. In this talk
we are mainly interested in the process-independent partonic factors $C_{ca}$ (see Eqs.~(11) and (14) in Ref.~\cite{Catani:2013tia}). Here the subscript $a$ ($a=q,{\bar q}, g$) denotes the type of initial-state colliding parton, while the subscript $c$  ($c=q,{\bar q}, g$) refers to the parton that produces the high-mass system through hard scattering. Within formulations~\cite{Luo:2019bmw, Gutierrez-Reyes:2019rug} of \qt\ resummation that are based on Soft Collinear Effective Theory (SCET) methods, the factors $C_{ca}$ are directly related to the so-called {\it matching coefficients} between transverse-momentum dependent (TMD) parton distributions and customary PDFs.

The quark collinear function $C_{qa}$ depends on the longitudinal-momentum fraction $z$ transferred in the collinear-radiation process, and it is computable
as a power series expansion in $\as$ as follows
\begin{align}
\label{eq:colq}
    C_{qa}\left(z;\as\right) = \delta_{qa}\delta(1-z) + \frac{\as}{\pi } \, C_{qa}^{(1)}\left(z\right)+ \sum_{m=2}^{\infty} \left( \frac{\as}{\pi } \right)^m C_{qa}^{(m)}\left(z\right)\,.
\end{align}
The gluon collinear function $C_{ga}^{\mu\nu}$ has a richer structure since it also depends on the Lorentz indices $\mu$ and $\nu$ of the gluon in the hard-scattering amplitude and its complex-conjugated amplitude, respectively that produces the high-mass system. The structure of the partonic tensor $C_{ga}^{\mu\nu}$ is \cite{Catani:2010pd}
\begin{align}
\label{eq:colg}
C_{ga}^{\mu\nu}\left(z;p_1,p_2,\mb{b};\as\right) = d^{\mu\nu}(p_1,p_2) \,C_{ga}(z;\as)+D^{\mu\nu}(p_1,p_2;\mathbf{b}) \,G_{ga}(z;\as)\,,
\end{align}
where $d^{\mu\nu}(p_1,p_2) = -g^{\mu\nu}+(p_1^{\mu}p_2^{\nu}+p_2^{\mu}p_1^{\nu})/(p_1p_2),\,\, D^{\mu\nu}(p_1,p_2;\mathbf{b})=d^{\mu\nu}(p_1,p_2)- 2 b^{\mu}b^{\nu}/\mb{b}^2$.
The light-like vectors $p_1^\mu$ and $p_2^\mu$ ($p_i^2=0, \;i=1,2$) in Eq.~(\ref{eq:colg}) denote the momenta of the initial-state colliding partons. In a reference frame in which the colliding hadrons are back-to-back, we can consider light-cone coordinates and we have $p_1^\mu = (p_1^{+},\mathbf{0_T},0)$ and $p_2^\mu = (0,\mathbf{0_T},p_2^{-})$. The momentum $b^\mu =  (0, \bb, 0)$ is the impact parameter vector in the four-dimensional notation. The gluon collinear functions $C_{ga}$ and $G_{ga}$ in Eq.~(\ref{eq:colg}) have the following perturbative expansions:
 \begin{align}
\label{azavgcolg}
   C_{ga}\left(z;\as\right) &= \delta_{ga}\delta(1-z) + \frac{\as}{\pi} \,C_{ga}^{(1)}\left(z\right)+ \sum_{m=2}^{\infty} 
\left(\frac{\as}{\pi}\right)^m \,
 C_{ga}^{(m)}\left(z\right)\,,
\\
\label{azcorcolg}
   G_{ga}\left(z;\as\right) &=  \frac{\as}{\pi} \,G_{ga}^{(1)}\left(z\right)+ \left(\frac{\as}{\pi}\right)^2 \,G_{ga}^{(2)}\left(z\right) +\sum_{m=3}^{\infty} \left(\frac{\as}{\pi}\right)^m \,G_{ga}^{(m)}\left(z\right)\,.
\end{align}
The structure of Eq.~(\ref{eq:colg}) is the consequence of collinear correlations \cite{Catani:2010pd} that are produced by the evolution of the colliding hadrons into gluon partonic states. In particular, the contribution of the tensor factor $D^{\mu \nu}$ in Eq.~(\ref{eq:colg}) leads to spin and azimuthal correlations 
\cite{Catani:2010pd, Catani:2017tuc} in the hard-scattering production of the observed high-mass system at small values of \qt. This contribution is sometimes denoted as the contribution of linearly-polarised gluons \cite{Mulders:2000sh} to TMD factorisation and \qt\ resummation. The size of the azimuthal correlations of collinear origin is controlled by the perturbative function $G_{ga}$ in Eq.~(\ref{azcorcolg}).

The azimuthally-uncorrelated quark and gluon collinear functions $C_{qa}$ and $C_{ga}$ in Eqs.~(\ref{eq:colq}) and 
(\ref{azavgcolg}) are known up to next-to-next-to-next-to-leading order~\cite{Luo:2019szz, Ebert:2020yqt, Luo:2020epw} in QCD perturbation theory,
namely up to ${\cal O}(\as^3)$. The azimuthally-correlated gluon collinear functions $G_{ga}(z;\as)$ in Eqs.~(\ref{eq:colg}) and (\ref{azcorcolg})
are instead known up to ${\cal O}(\as^2)$. The first-order coefficients $G_{ga}^{(1)}$ were computed in Ref.~\cite{Catani:2010pd}. The second-order terms $G_{ga}^{(2)}(z)$ have been obtained more recently
in Refs.~\cite{Luo:2019bmw} and 
\cite{Gutierrez-Reyes:2019rug}. {\it Our independent computation} of $G_{ga}^{(2)}(z)$ confirms the results of Refs.~\cite{Luo:2019bmw, Gutierrez-Reyes:2019rug}.

The \qt\ resummation formalism can be extended to processes that are related by {\it kinematical crossing}
to the hadroproduction processes. The corresponding resummation formulae
are analogous to that for hadron--hadron collisions and they involve a main difference through the replacement of the PDFs with the parton fragmentation functions (PFFs) of the triggered hadrons in the final state.
In the resummation formulae the PFFs are convoluted with computable perturbative functions that embody the effect of QCD radiation
collinear to the final-state partons. These perturbative functions for the time-like (TL) collinear evolution have the same structure as the initial-state collinear functions
$C_{qa}, C_{ga}$ and $G_{ga}$ in Eqs.~(\ref{eq:colq}, \ref{eq:colg}, \ref{azavgcolg}) and (\ref{azcorcolg}), and they are denoted
by $C_{qa}^{TL},\, C_{ga}^{TL}$, and $G_{ga}^{TL}$ in this talk.
\section{Details of our computational method: differential and TMD collinear functions}
\label{method}
In this talk, we compute the collinear functions introduced in Sec.~\ref{intro} starting from the evaluation of QCD scattering amplitudes.
At the bare level, the computation exhibits ultraviolet (UV) and infrared (IR) divergences. We regularise both the divergences by working in
$d=4 - 2\ep$ space-time dimensions. In particular, we use the conventional dimensional regularisation (CDR) scheme,
in which on-shell gluons have $d-2$ physical states of spin polarisations and on-shell massless quarks (or antiquarks) have 2 spin polarisation states. The dimensional regularisation scale is denoted by $\mu_0$.

QCD scattering amplitudes are singular in the kinematical configurations in which two or more momenta of their external massless
partons become collinear. The singular behaviour in the collinear limit is described by a factorisation formula that has a process-independent structure. Following Ref.~\cite{Catani:2011st}, we write the collinear factorisation formula in its most general form as follows
\begin{equation}
\label{colfact}
|  {\cal M}\left( \{ q_i \}; k_1, \dots, k_N \right) |^2 =
\langle  {\cal M}\bigl( \{ q_i \}; {\tilde k} \bigr) | 
\;\bcp\bigl( \{ q_i \};  k_1, \dots, k_N ; n\bigr) \;
|  {\cal M}\bigl( \{ q_i \}; {\tilde k} \bigr) \rangle + \dots \;,
\end{equation}
where the dots on the right-hand side denote non-singular terms in the collinear limit.
Here, $\cal M$ denotes the on-shell scattering amplitude in colour+spin space notation of a generic hard-scattering process, and
$|{\cal M}|^2$ is the corresponding squared amplitude.
In Eq.~(\ref{colfact}) we are considering the limit in which the momenta $k_1, \dots, k_N$ of $N$ external massless QCD partons become collinear. The momenta of the other external partons of $\cal M$
are $q_1,q_2,\dots$ and their dependence through momenta and other quantum numbers is denoted by $\{ q_i \}$. The singular behaviour in the collinear limit is embodied in the factor
$\bcp$, while ${\cal M}\bigl( \{ q_i \}; {\tilde k} \bigr)$ denotes the reduced scattering amplitude that is obtained
from ${\cal M}\left( \{ q_i \}; k_1, \dots, k_N \right)$ by replacing the $N$ collinear partons with a single parent parton with momentum
${\tilde k} $.

In Eq.~(\ref{colfact}), $\tilde k$ and $k_1, \dots, k_N$ are the {\it outgoing} 
momenta of the corresponding external partons.
The scattering amplitude $\cal M$ (and the kernel $\bcp$) is evaluated in different physical kinematical regions depending on the sign
of the `energies' of the outgoing momenta. If the energies of $k_1, \dots, k_N$ are all positive, we are dealing
with the TL collinear region. Otherwise, we are considering the space-like (SL) collinear region. The distinction between TL and SL collinear regions is,
in general, very relevant. Indeed, in the case of the TL collinear region the splitting kernel $\bcp$ has the relevant property of being completely
process independent: it does not depend on the momenta and quantum numbers of the non-collinear partons in $\cal M$.
This property of strict collinear factorisation is instead violated in the SL collinear regions \cite{Catani:2011st}. The collinear splitting kernels $\bcp$ at ${\cal O}(\as)$ are well known and at ${\cal O}(\as^2)$, they are fully known 
\cite{Campbell:1997hg, Catani:1999ss, Bern:1999ry, Kosower:1999rx, Sborlini:2013jba}
for both the TL and SL collinear regions \cite{Catani:2011st}. 

Considering the TL region, the splitting kernels $\bcp$ are process independent and proportional to the unit matrix in colour space. However, the dependence of $\bcp_{c \to a_1 \dots a_N}$ (where $a_i$ denotes the flavour of the collinear parton with momentum $k_i$, and $c$ is the flavour of the parent parton) on the spin of the parent collinear parton $c$ can be instead non-trivial \cite{Catani:1999ss} and it is different for the cases $c=q,{\bar q}$ and $c=g$. In the case of collinear splitting of a quark or antiquark, spin correlations are completely absent while they are present in the collinear splitting of a gluon through a rank-2 tensor.

The tensor dependence of $\cp^{\mu \nu}_{g \to a_1 \cdots a_N}$
is due to terms that are proportional to either the metric tensor $g^{\mu \nu}$
or to quadratic terms of the type $k_{i \,T}^\mu k_{j \,T}^\nu$, where
$k_{i \,T}^\mu$ is the transverse momentum of the $i$-th collinear parton with respect to the collinear direction.
The remaining dependence of $\cp_{c \to a_1 \cdots a_N}$ ($c=q,\bar q,g$)
is due to scalar functions of the collinear momenta $k_1, \dots, k_N$. These functions are the sub-energies $s_{ij}= 2 k_i k_j$ and the ratios $x_i/x_j$
of the longitudinal-momentum fractions $x_i$ and $x_j$ of the momenta  $k_i$ and
$k_j$. The most general definition \cite{Catani:1999ss} of the longitudinal-momentum fractions is obtained by introducing an auxiliary reference
vector $n^\mu$ that is far away from the collinear direction. Then the collinear splitting kernels depend on the ratios $x_i/x_j$ that are defined as $x_i/x_j = nk_i/nk_j$. In the literature the reference vector $n^\mu$ is usually chosen to be a {\it light-like} (i.e., $n^2=0$)
vector.
However, we emphasise that one can also set $n^2\neq 0$. In this talk we introduce and use a {\it time-like} (i.e., $n^2 > 0$) auxiliary vector $n^\mu$. Note that we do not modify any formal expression of the splitting kernels $\bcp$ in the literature. Simply, we use the freedom of arbitrarily choosing $n^2 \geq 0$ in the collinear limit.

After our discussion of the structure of Eq.~(\ref{colfact}) in the TL collinear region, we define differential TL collinear functions ${\cf}^{\rm TL}_{ca}$
($c,a=g, q, \bar q$) as follows~\cite{Catani:2022sgr}. We consider the production of a parton of flavour $a$ and momentum $p^\mu$ in the physical final state and we fully integrate over the accompanying collinear radiation by keeping its total momentum $k$ fixed. If the parent collinear parton $c$ is a gluon, we have to take into account
the spin correlations, and the precise definition of the collinear function $\cf^{\rm{TL} \, \mu \nu}_{ga}$ is
\begin{align}
\label{gcolfun}
    \cf^{\rm{TL} \,\mu \nu}_{ga}(p,k;n) &= \sum_{N=2}^{+\infty}
\left[\prod_{m=1}^{N-1}\int \frac{d^d k_m}{(2\pi)^{d-1}} \delta_{+}\left(k_m^2\right)\right] \;\delta^{(d)}\!\!\left(k-\sum_{i=1}^{N-1} k_i\right) \nn \\
& \times \sum_{a_1,\dots,a_{N-1}}
\frac{{\widetilde \cp}_{g\rightarrow a_1 \dots a_N}^{\mu \nu}\left(k_1,\dots,k_N;n\right)}{{\rm SF}\left(a_1,\dots,a_{N-1}\right)}{\Bigg |}_{\substack{k_N = \,p \\
a_N = \,a}} \;\;,
\end{align}
where ${\rm SF}\left(a_1,\dots,a_{N-1}\right)$ is the Bose symmetry factor for the identical particles in the final state. The parton momentum $p^\mu$ precisely specifies the collinear direction, and we can use a light-cone reference frame where $p^\mu= (p^+,\mathbf{0_T},0)$ with
$p^+ > 0$. In this frame we have $k_i^\mu = (k_i^+,\mathbf{k_{i\, T}},k_i^-)$
and $k^\mu = (k^+,\mathbf{k_{T}},k^-)$. The auxiliary TL vector $n$ has coordinates $n^\mu=( n^+, \mathbf{0_T}, n^-)$, with $n^2=2n^+n^- > 0$. The case of a light-like vector $n^\mu$ is obtained by setting $n^+ = 0$. The gluonic kernel ${\widetilde \cp}^{\mu \nu}$ in Eq.~(\ref{gcolfun}) is given by ${\widetilde \cp}_{g\rightarrow a_1 \dots a_N}^{\mu \nu}\left(k_1,\dots,k_N;n\right) =
d^\mu_{\;\; \mu^\prime}(p;n) \;{\cp}_{g\rightarrow a_1 \dots a_N}^{\mu^\prime \nu^\prime}\left(k_1,\dots,k_N;n\right) \;d_{\nu^\prime}^{\;\;\; \nu}(p;n),$
where the spin polarization tensor, $\dumunu = -g^{\mu\nu}+(p^{\mu}n^{\nu}+n^{\mu}p^{\nu})/(np)-(n^2p^{\mu}p^{\nu})/(np)^2\; .$
The use of ${\widetilde \cp}^{\mu \nu}$ in Eq.~(\ref{gcolfun}) removes purely longitudinal terms, proportional to $p^\mu$ or $p^\nu$, from 
$\cf^{\rm{TL} \,\mu \nu}_{ga}$.
The function $\cf^{\rm{TL} \,\mu \nu}$ depends on the vectors $p,k,n$ and it is orthogonal to both $p$ and $n$. Therefore it has the following decomposition in tensor structures:
\begin{equation}
\label{fincor}
\cf^{\rm{TL} \,\mu \nu}_{ga}(p,k;n) = 
d^{\mu \nu}(p;n) \;\cf^{\rm{TL}}_{ga,\,{\rm az.in.}}(p,k;n)
+ D^{\mu\nu}(p,n;\mb{k_T},\ep) \;\cf^{\rm{TL}}_{ga,\,{\rm corr.}}(p,k;n) \;\;,
\end{equation}
where $D^{\mu\nu}(p,n;\mb{k_T},\ep) = d^{\mu\nu}(p;n)-(d-2)k_T^{\mu}k_T^{\nu}/\mb{k_T}^2$. The tensor $D^{\mu\nu}$ in Eq.~(\ref{fincor}) leads to correlations with respect to the azimuthal angle of the transverse-momentum vector $\mb{k_T}$. The scalar functions $\cf^{\rm{TL}}_{ga,\,{\rm az.in.}}$ and $\cf^{\rm{TL}}_{ga,\,{\rm corr.}}$ control the size of the azimuthal un-correlated and correlated contributions to $\cf^{\rm{TL} \,\mu \nu}_{ga}$, respectively. The tensors in Eq.~(\ref{fincor}) are the $d$-dimensional generalisation of those in Eq.~(\ref{eq:colg}) and they fulfil following relations: $\ddmunu D^{\mu\nu}(p,n;\mb{k_T},\ep)=0,\,\ddmunu \,\dumunu= d-2,\,D_{\mu\nu}(p,n;\mb{k_T},\ep)D^{\mu\nu}(p,n;\mb{k_T},\ep)=(d-2)(d-3),\,$ with $d-2= 2-2\eps$ and $d-3=1-2\eps$. The scalar functions $\cf^{\rm{TL}}_{ga,\,{\rm az.in.}}$ and $\cf^{\rm{TL}}_{ga,\,{\rm corr.}}$ can be easily and directly expressed in terms of the collinear splitting kernels ${\cp}_{g\rightarrow a_1 \dots a_N}^{\mu\nu}$ using the above relations. The TL collinear function $\cf^{\rm{TL}}_{ca}(p,k;n)$ $(c=q,\bar q)$ of a parent collinear fermion $c$ is defined analogously to the gluon collinear function $\cf^{\rm{TL}}_{ga}(p,k;n)$ by simply performing the replacements 
$\cf^{\rm{TL} \,\mu \nu}_{ga} \to \cf^{\rm{TL}}_{ca}$ and
${\widetilde \cp}_{g\rightarrow a_1 \dots a_N} \to 
{\cp}_{c\rightarrow a_1 \dots a_N}$ in Eq.~(\ref{gcolfun}).

The SL collinear splitting kernels $\bcp$ in the factorisation formula (\ref{colfact})
are, in general, process dependent and, in particular, they can depend on the colour indices and momenta of the non-collinear partons and on the colour indices of the parent collinear parton $c$
in the hard-scattering process. The general extension $\bcf_{ca}$ of the TL collinear function in 
Eqs.~(\ref{gcolfun}) to the SL collinear regions 
is as follows
\begin{align}
\label{slcolfun}
    \bcf_{ca}(\{q_i\};p,k;n) &= \sum_{N=2}^{+\infty}
\left[\prod_{m=1}^{N-1}\int \frac{d^d k_m}{(2\pi)^{d-1}} \delta_{+}\left(k_m^2\right)\right] \;\delta^{(d)}\!\!\left(k-\sum_{i=1}^{N-1} k_i\right) \\
& \times \sum_{a_1,\dots,a_{N-1}}
\frac{{\widetilde \bcp}_{{\bar c}\rightarrow a_1 \dots a_N}\left(\{q_i\};k_1,\dots,k_N;n\right)}{{\rm SF}\left(a_1,\dots,a_{N-1}\right)}{\Bigg |}_{\substack{k_N=\,- p \\
a_N = \, {\bar a}}} \;\; \frac{\mathcal{N}_c(\eps)}{\mathcal{N}_a(\eps)}\;\;, \quad \;\;\; c=g,q,{\bar q}\nn \;,
\end{align}
where $a$ denotes the parton with momentum $p^\mu$ that collides in the physical initial state and $c$ refers to the incoming parton of the hard-scattering process
after the radiation of the final-state collinear partons with total momentum 
$k^\mu$. The function ${\mathcal{N}_a(\eps)}$ in  
Eq.~(\ref{slcolfun}) is given by $\mathcal{N}_a(\eps) = (-1)^{2S_a}n_s(a,\eps)n_c(a)$,
where $S_a$ denotes the spin, 
$n_s(a,\eps)$ denotes the number of spin polarisation states, and
$n_c(a)$ denotes the number of colours of the parton $a$. Therefore, we have 
$\mathcal{N}_{q}(\eps) = \mathcal{N}_{{\bar q}}(\eps)= -2 N_c$ 
and $\mathcal{N}_g(\eps) = 2(1-\eps)(N_c^2-1)$.

Considering perturbative contributions at ${\cal O}(\as)$ and ${\cal O}(\as^2)$,
the SL collinear kernels $\bcp$ and functions $\bcf$ are process independent 
and proportional to the unit matrix in the colour space of the hard-scattering partons.
Therefore, we can factorise such overall (and trivial) colour space dependence
in both sides of Eq.~(\ref{slcolfun}), and we can simply deal with $c$-number 
SL collinear functions, analogously to the TL collinear case. Such SL collinear
functions are denoted as  $\cf^{\, \mu \nu}_{ga}(p,k;n)$, $\cf_{ga,\,{\rm az.in.}}(p,k;n)$,
$\cf_{ga,\,{\rm corr.}}(p,k;n)$, and $\cf_{ca}(p,k;n) \;(c=q,{\bar q})$.

The TL and SL differential collinear functions can be used to define inclusive functions that are directly related to the perturbative computation and resummation
of large logarithmic contributions to hard-scattering observables. In the following we define TMD collinear functions~\cite{Catani:2022sgr} that lead to the resummation coefficients which we have discussed in the Sec.~\ref{intro}. We first consider the TL collinear region. We use the gluon collinear function $\cf^{\rm{TL} \,\mu \nu}_{ga}(p,k;n)$ 
of Eq.~(\ref{gcolfun}) and we define the gluon TMD function  $F^{\rm{TL} \,\mu \nu}_{ga}$ by integrating over
the radiated collinear momentum $k$ 
as follows
\begin{align}
\label{tlgtmd}
 F^{\rm{TL} \,\mu \nu}_{ga}(z;p/z,\mb{q_T};n) &= \delta(1-z) \;\delta^{(d-2)}(\mb{q_T})
\;\, \delta_{ga} \;\, d^{\mu \nu}(p;n) \nn \\
&+ 
\int d^dk \;\delta^{(d-2)}(\mb{k_T}+\mb{q_T}) 
\;\,\delta\!\left(\frac{k^{+}}{p^{+}}-\frac{1-z}{z}\right)
\;\cf^{\rm{TL} \,\mu \nu}_{ga}(p,k;n)
\;\;.
\end{align}
The TMD function  $F^{\rm{TL} \, \mu \nu}_{ga}$ describes the inclusive perturbative fragmentation of a gluon into a parton $a$. Similar to Eq.~(\ref{fincor}), the Eq.~(\ref{tlgtmd}) can be further decomposed into $F^{\rm{TL}}_{ga,\,{\rm az.in.}}$ and $F^{\rm{TL}}_{ga,\,{\rm corr.}}$ components. The quark (or antiquark) TMD function $F^{\rm{TL}}_{ca}$ ($c=q, {\bar q}$) can be defined analogously to the gluon TMD function 
in Eq.~(\ref{tlgtmd}).
We note that $F^{\rm{TL} }_{ga,\,{\rm az.in.}}, F^{\rm{TL} }_{ga,\,{\rm corr.}}$, and
$F^{\rm{TL}}_{ca}$ are scalar functions that depend on $z$ and the vectors $p^\mu, n^\mu, \mb{q_T}$ through the functional form 
$z, \mb{q_T}^2$ and $n^2 \mb{q_T}^2/(2np/z)^2$.

The general SL TMD function ${\bf F}_{ca}$ is obtained by analogy with the TL function in Eq.~(\ref{tlgtmd}) and by taking into account that the SL collinear function $\bcf_{ca}$ in Eq.~(\ref{slcolfun}) is, in general, process
dependent. The explicit definition of ${\bf F}_{ca}$ is
\begin{align}
\label{sltmdgen}
 {\bf F}_{ca}(\{q_i\};z;zp,\mb{q_T};n) &= \mathbf{1} \;
\delta(1-z) \;\delta^{(d-2)}(\mb{q_T}) \;\, \delta_{ca}
\nn \\
&+ z
\int d^dk \;\,\delta^{(d-2)}(\mb{k_T}+\mb{q_T}) 
\;\,\delta\!\left(\frac{k^{+}}{p^{+}}-1+ z\right)
\;\bcf_{ca}(\{q_i\};p,k;n)
\;.
\end{align}
The TMD function ${\bf F}_{ca}$ is a process-dependent operator in colour$+$spin space. However, up to $\mathcal{O}(\alpha_{\rm S}^2)$ considered in this talk, the TMD function ${\bf F}_{ca}$ is process independent following the process independence of $\bcf_{ca}$.
\section{Perturbative results}
\label{results}
In this section we discuss the perturbative calculation of the SL collinear functions introduced in Sec.~\ref{method}.
We consider explicit computations up to ${\cal O}(\as^2)$ and, hence, we simply refer to the process-independent $c$-number functions. We left the computation of TL collinear functions to our original article. The perturbative expansion of the collinear functions can be written as follows
\begin{equation}
\label{fexp12}
\cf(p,k;n) = \cf^{(1R)}(p,k;n) + 
\left[ \; \cf^{(2R)}(p,k;n) + \cf^{(1R1V)}(p,k;n) \; \right] 
+ {\cal O}(\as^3) \;\;.
\end{equation}
The notation in Eq.~(\ref{fexp12}) applies to any specific collinear function and, therefore, we have not explicitly denoted the corresponding subscripts and superscripts in $\cf$. The contribution to $\cf$ at ${\cal O}(\as)$ is due to
$\cf^{(1R)}$, which corresponds to single real emission in the final state at the tree-level. The contributions to $\cf$ at
${\cal O}(\as^2)$ are $\cf^{(2R)}$ (double real emission at the tree level)
and $\cf^{(1R1V)}$ (single real emission with one-loop virtual corrections).

We express the contributions in Eq.~(\ref{fexp12}) in terms of the unrenormalised (bare) QCD coupling $\alpha^u_{\rm S}$, which is related to the renormalised coupling $\as(\mu_R^2)$ in the
$\overline{{\rm MS}}$ renormalisation scheme as follows:
\begin{align}
\label{eq:alphasren}
    \alpha^u_{\rm S} \,\mu_0^{2\eps} \,S_{\eps} = \as(\mu_R^2) \,\mu_R^{2\eps}\left[1-\frac{\as(\mu_R^2)}{\pi}\frac{\beta_0}{\eps} +\mc{O}\left(\as^2(\mu_R^2)\right)   \right]\,,
\end{align}
where $\beta_0= (11 C_A - 2 N_f)/12$ and $N_f$ is the number of massless-quark flavours. 
The $d$-dimensional spherical factor $S_{\eps}$ is
$S_{\eps}=\left(4 \pi \,e^{-\gamma_E} \right)^{\eps}$ and $\gamma_E$ is the Euler number
($\gamma_E=0.5772\dots$).

In the case of the azimuthally-independent functions 
$\cf^{(1R)}_{ca, \,\text{az.in.}}$, the collinear kernels are proportional to ${\widehat P}_{ca}(x; \ep)$~\cite{Catani:1999ss}, which are the $d$-dimensional real emission contributions to the Altarelli--Parisi splitting functions for the leading order evolution of the PDFs. We have
\begin{equation}
\label{f1R}
\cf^{(1R)}_{ca, \,\text{az.in.}}(p,k;n) = \frac{\alpha^u_{\rm S} \,\mu_0^{2\eps}
\,S_{\eps}}{\pi} \;\frac{e^{\eps\gamma_E}}{\pi^{1-\eps}} 
\;\frac{\delta_{+}(k^2)}{pk} \;\frac{1}{z_n} \;{\widehat P}_{ca}(z_n; \ep) \;\;,
\;\;\; \quad c=g,q,{\bar q} \;\;,
\end{equation}
where we have introduced the notation 
$\cf^{(1R)}_{ca} \equiv \cf^{(1R)}_{ca, \,\text{az.in.}}$ ($c=q,{\bar q}$) for the 
collinear functions in the quark and antiquark partonic channels.
The azimuthal-correlation contribution in the gluon channel 
is
\begin{equation}
\label{fgacor}
\mathcal{F}^{(1R)}_{ga, \,\text{corr.}}(p,k;n) =- \;\frac{\alpha^u_{\rm S} \,\mu_0^{2\eps} \,S_{\eps}}{\pi} \;\frac{e^{\eps\gamma_E}}{\pi^{1-\eps}} 
\;\frac{\delta_{+}(k^2)}{pk} \;C_a \;\frac{1-z_n}{z_n^2} 
\;\;,
\end{equation}
where $C_a$ is the Casimir colour coefficient of the parton $a=q,{\bar q},g$. 
The expressions of $\mathcal{F}^{(1R)}$ in Eqs.~(\ref{f1R}) and (\ref{fgacor})
depend on the auxiliary vector $n^\mu$ through the variable $z_n$, $z_n = n (p-k)/np$.
In the exact collinear limit (i.e., $k^- = 0$) the parent hard-scattering parton $c$ in $\mathcal{F}^{(1R)}_{ca}$ carries the momentum $z_n p^\mu$, independently of the value of $n^2$.

The SL TMD functions $F_{ca}$ are obtained from the differential collinear functions
$\cf_{ca}$ by using Eq.~(\ref{sltmdgen}). The terms $\cf_{ca}^{(1R)}$, 
$\cf_{ca}^{(2R)}$, and $\cf_{ca}^{(1R1V)}$ in Eq.~(\ref{fexp12}) produce corresponding contributions to the TMD functions that are denoted as
$F_{ca}^{(1R)}$, $F_{ca}^{(2R)}$, and $F_{ca}^{(1R1V)}$, respectively.

We immediately discuss the dependence on the auxiliary vector $n^\mu$, which affects 
$F_{ca}^{(1R)}$ through the variable $z_n$. The key point regards the effect of the singular contribution of ${\widehat P}_{ca}(z_n;\ep)$ to $\cf_{ca}^{(1R)}$ and, hence, to $F_{ca}^{(1R)}$.
Such contribution is proportional to the following factor:
\begin{equation}
\label{znsing}
\frac{1}{1-z_n} = \frac{np}{nk} = \frac{p^+}{k^+ + \frac{n^2}{2 np}  \frac{k^-}{n^-} p^+}
= \frac{1}{1-z +\frac{n^2 \mathbf{q_T}^2}{(1-z) (2 np)^2}} \;\;,
\end{equation}
where in the last equality we have implemented the kinematics of the TMD collinear function at ${\cal O}(\as)$ (i.e., $k^2=0,\,k^+=(1-z)p^+,\,\mb{k_T}= - \mb{q_T}$).
Setting $n^2=0$, the factor in Eq.~(\ref{znsing}) becomes $(1-z)^{-1}$ and, therefore, it is divergent (and not integrable over $z$) at $z=1$. Correspondingly, the first-order contributions $F_{ca}^{(1R)}$ to the TMD collinear functions are divergent. Such divergences, which are known as 
rapidity divergences in the literature, are a general feature of SCET formulations of TMD functions, and they can be treated by introducing appropriate regularisation 
procedures. In our computations
of the TMD functions we use $n^2 > 0$, thus
avoiding rapidity divergences. Indeed,
setting $\lambda = n^2 \mb{q_T}^2/{(2pn)^2}$ in Eq.~(\ref{znsing}), we can use the following relation:
\begin{align}
\label{tmdplus}
\frac{1}{1-z +\frac{\lambda}{1-z}} &= \left( \frac{1-z}{(1-z)^2 +\lambda} \right)_+
+\delta(1-z) \int_0^1 dz^\prime \frac{1-z^\prime}{(1-z^\prime)^2 +\lambda} \nn \\
&= \frac{1}{2} \;\ln\left(\frac{1}{\lambda}\right) \;\delta(1-z) 
+ \left( \frac{1}{1-z} \right)_+ + {\cal O}({\sqrt \lambda})
\;\;,
\end{align}
where the symbol $\bigl( f(z) \bigr)_+$ denotes the customary `plus-distribution'
of the function $f(z)$ with respect to the variable $z$.
The term of ${\cal O}({\sqrt \lambda}) \sim {\cal O}(\mb{q_T})$ in Eq.~(\ref{tmdplus})
smoothly vanishes in the limit $\mb{q_T} \to 0$ and, therefore, it can be neglected in the computation of $F_{ca}^{(1R)}$. We can similarly neglect other smooth terms 
in the limit $\mb{q_T} \to 0$ by using $z_n= z + {\cal O}(\mb{q_T}^2)$ in the remaining
$z_n$ dependence of $F_{ca}^{(1R)}$.

We introduce the Fourier transformation of the TMD collinear function $F_{ca}$ to the purpose of having a more direct relation with the discussion in Sec.~\ref{intro}.
The Fourier transformation ${\widetilde F}_{ca}$ in $\mb{b}$ space of the TMD collinear function $F_{ca}$ for the quark and antiquark partonic channels is
\begin{equation}
\label{tmdqb}
{\widetilde F}_{ca}\!\left( z;\frac{\mb{b}^2}{b_0^2},\frac{n^2 b_0^2}{(2zpn)^2 \,\mb{b}^2}\right)
\equiv \int d^{d-2}\mb{q_T}\;e^{-i\mb{b}.\mb{q_T}} \;{F}_{ca}\!\left(z;\mb{q_T}^2,\frac{n^2 \mb{q_T}^2}{(2zpn)^2}\right) \;,
\end{equation}
where the impact parameter $\mb{b}$ is a $(d-2)$-dimensional vector. The numerical coefficient, $b_0=2 e^{-\gamma_E}$.
We note that $n^\mu$ dependence of ${\widetilde F}_{ca}$ occurs through
the variable  $\tilde{\lambda}= n^2 (b_0^2/\mb{b}^2)/(2zpn)^2$. Analogously, we can introduce the Fourier transformation in the gluon channel. 

In general, the perturbative computation at ${\cal O}(\as^n)$
of the TMD functions in $\mb{b}$ space leads to divergent pole terms $1/\ep^m$ with
$1 \leq m \leq 2n$. These divergences are of UV and IR origin. The UV divergences are removed by using Eq.~(\ref{eq:alphasren}). The IR divergences are then factorisable. The TMD functions in $\mb{b}$ space fulfil the following IR factorisation formulae~\cite{Catani:2022sgr}:
\begin{align}
&{\widetilde F}_{ca,\,{\rm az.in.}}\!\!\left(z;\frac{\mb{b}^2}{b_0^2},
\frac{n^2 b_0^2}{(2zpn)^2 \,\mb{b}^2}\right) = 
Z_c\!\left(\as(b_0^2/\mb{b}^2), \frac{n^2 b_0^2}{(2zpn)^2 \,\mb{b}^2}\right) \nn \\
& \quad \quad \times \sum_b \int_z^1 \frac{dx}{x} \;
{\widetilde C}_{cb}\!\left(x;\as(b_0^2/\mb{b}^2), \ep, \frac{n^2 b_0^2}{(2zpn)^2 \,\mb{b}^2}\right) \;\; {\widetilde \Gamma}_{ba}(z/x;b_0^2/\mb{b}^2) \;\;,
\label{IRfactav}
\end{align}
\begin{align}
&{\widetilde F}_{ga,\,{\rm corr.}}\!\!\left(z;\frac{\mb{b}^2}{b_0^2},
\frac{n^2 b_0^2}{(2zpn)^2 \,\mb{b}^2}\right) = 
Z_g\!\left(\as(b_0^2/\mb{b}^2), \frac{n^2 b_0^2}{(2zpn)^2 \,\mb{b}^2}\right) \nn \\
& \quad \quad \times \sum_b \int_z^1 \frac{dx}{x} \;
{\widetilde G}_{gb}\!\left(x;\as(b_0^2/\mb{b}^2), \ep, \frac{n^2 b_0^2}{(2zpn)^2 \,\mb{b}^2}\right) \;\; {\widetilde \Gamma}_{ba}(z/x;b_0^2/\mb{b}^2) \;\;.
\label{IRfactcor}
\end{align}
Note that in the right-hand side of Eqs.~(\ref{IRfactav}) and (\ref{IRfactcor}) we use the renormalization scale
$\mu_R^2=b_0^2/\mb{b}^2$. Therefore, the various functions $Z_c, {\widetilde \Gamma}_{ba},
{\widetilde C}_{cb}$ and ${\widetilde G}_{cb}$ depend on $\as(b_0^2/\mb{b}^2)$. The factor ${\widetilde \Gamma}_{ba}(x;\mu_F^2)$ is the customary collinear-divergent function
that defines the scale-dependent PDF $f_b(z;\mu_F^2)$ in the $\overline{{\rm MS}}$
factorisation scheme. After factorisation of the collinear $\ep$ poles,
the $\mb{b}$ space TMD functions still contain IR divergences that are factorisable in the
perturbative functions $Z_c$ of 
Eqs.~(\ref{IRfactav}) and (\ref{IRfactcor}). The functions ${\widetilde C}_{cb}$ and ${\widetilde G}_{gb}$ are then finite and independent of $n^2$ 
(i.e., $n^2 (b_0^2/\mb{b}^2)/(2zpn)^2$) in the limit $\ep \to 0$, order-by-order in the perturbative expansion in powers of $\as(b_0^2/\mb{b}^2)$.

The IR finite function ${\widetilde C}_{ca}$ of the azimuthal-independent component
of the TMD function in Eq.~(\ref{IRfactav}) has the following perturbative expansion:
\begin{equation}
{\widetilde C}_{ca}(z;\as,\ep,{\widetilde \lambda}) = \delta_{ca} \;\delta(1-z) +
\frac{\as}{\pi} \;{\widetilde C}_{ca}^{(1)}(z;\ep,{\widetilde \lambda})
+ \left(\frac{\as}{\pi}\right)^2 \;{\widetilde C}_{ca}^{(2)}(z;\ep,{\widetilde \lambda})
+ {\cal O}(\as^3)\;.
\label{ctilde}
\end{equation}
The limit $\ep \to 0$ in Eq. (\ref{ctilde}) gives the collinear functions in Eqs.~(\ref{eq:colq}) and (\ref{azavgcolg}),
namely ${\widetilde C}_{ca}^{(m)}(z;\ep=0,{\widetilde \lambda}) = C_{ca}^{(m)}(z)
\;\; (c=q,{\bar q},g)$. The IR finite function ${\widetilde G}_{ga}$ in Eq.~(\ref{IRfactcor}) has the following perturbative expansion:
\begin{equation}
{\widetilde G}_{ga}(z;\as,\ep,{\widetilde \lambda}) = 
\frac{\as}{\pi} \;{\widetilde G}_{ga}^{(1)}(z;\ep,{\widetilde \lambda})
+ \left(\frac{\as}{\pi}\right)^2 \;{\widetilde G}_{ga}^{(2)}(z;\ep,{\widetilde \lambda})
+ {\cal O}(\as^3) \;\;. 
\label{gtilexp}
\end{equation}
In the four-dimensional limit $\ep \to 0$, ${\widetilde G}_{ga}$ gives the 
transverse-momentum resummation function $G_{ga}$ in 
Eqs.~(\ref{eq:colg}) and (\ref{azcorcolg}),
and specifically we have ${\widetilde G}_{ga}^{(m)}(z;\ep=0,{\widetilde \lambda}) = 
G_{ga}^{(m)}(z)$. We find the following results for $G_{ga}^{(m)}(z)$ at $\mathcal{O}(\alpha_{\rm S}^2)$:
\begin{align}
\label{ggcorr2}
G^{(2)}_{gg}(z) &= C_A^2\bigg\{-\frac{37}{36z}+\frac{31}{18}-\frac{13z}{12}+\frac{11z^2}{36}-\ln(z)\left[\frac{1}{z}+\frac{19}{12}\right]+\frac{\ln^2(z)}{2}+\frac{1-z}{z}\left[\text{Li}_2(z)-\frac{\pi^2}{6}\right]\bigg\} \nn \\
& + C_F N_f\left\{\frac{(1-z)^3}{2z}-\frac{1}{4}\ln^2(z)\right\}+C_A N_f\left\{-\frac{17}{36z}+\frac{4}{9}+\frac{z}{12}+\frac{z^2}{36}-\frac{1}{6}\ln(z)\right\} \nn \\
&
- h^{(1)}_g \,C_A \,\frac{1-z}{z} \;\;,\\
\label{gqcorr2}
G^{(2)}_{gq}(z) &= C_F^2\left\{-\frac{1-z}{2}+\frac{5}{4}\ln(z)-\frac{1}{4}\ln^2(z)-\frac{1-z}{2z}\bigg[\ln(1-z)+\ln^2(1-z)\bigg] \right\}
\nn\\
&+C_F N_f\left\{-\frac{1-z}{3z}\left[\frac{2}{3}+\ln(1-z)\right] \right\}+ C_A C_F\bigg\{-\frac{11}{18z}+\frac{10}{9}-\frac{z}{2}-\ln(z)\left[\frac{1}{z}+\frac{5}{2}\right]
\nn\\
&+\frac{1}{2}\ln^2(z)+\frac{1-z}{z}\left[\frac{5}{6}\ln(1-z)+\frac{1}{2}\ln^2(1-z)+\text{Li}_2(z)-\frac{\pi^2}{6}\right] \bigg\}
- h^{(1)}_g \,C_F \,\frac{1-z}{z} \;\;,
\end{align}
and $G^{(2)}_{g{\bar q}}(z)=G^{(2)}_{gq}(z)$. In the above,  $h_g^{(1)}$ is a scheme dependent (see Ref.~\cite{Catani:2013tia} for more details) coefficient. Our results in Eqs.~(\ref{ggcorr2}) and (\ref{gqcorr2}) are in full agreement with the literature~\cite{Luo:2019bmw, Gutierrez-Reyes:2019rug} and hence, provide a non-trivial check on our theoretical framework and explicit computations.
\section{Conclusion}
In this talk we have considered the computation of collinear contributions to the transverse momentum resummation up to $\mathcal{O}(\alpha_{\rm S}^2)$. Starting from the factorisation structure of scattering amplitudes in the collinear limit, we have introduced differential collinear functions that have a process independent structure. These collinear functions upon integration over the appropriate observable-dependent phase space gives the logarithmically-enhanced contributions to the corresponding observable. Through our formalism we have shown how these collinear functions can be defined without introducing what is known as rapidity divergences in the literature and presented explicit results for the azimuthally-correlated contributions to the TMD collinear function at $\mathcal{O}(\alpha_{\rm S}^2)$.
\section*{Acknowledgement}
PKD is grateful to Stefano Catani for many insightful discussions.
\bibliography{main}
\bibliographystyle{JHEP}
\end{document}